# Strong Modulation of Optical Properties in Rippled 2D GaSe *via* Strain Engineering

**David Maeso[1], Sahar Pakdel[1,2], Hernán Santos[1], Nicolás Agraït[1,3,4,5], Juan José Palacios[1,4,5], Elsa Prada[1,4,5*] and Gabino Rubio-Bollinger[1,4,5*]**

[1]Departamento de Física de la Materia Condensada, Universidad Autónoma de Madrid, E-28049 Madrid, Spain.
[2]School of Electrical and Computer Engineering, University College of Engineering, University of Tehran, Tehran 14395-515, Iran.
[3]Instituto Madrileño de Estudios Avanzados en Nanociencia (IMDEA-Nanociencia), E-28049 Madrid, Spain., Spain.
[4]Condensed Matter Physics Center (IFIMAC), Universidad Autónoma de Madrid, E-28049 Madrid
[5]Instituto Nicolás Cabrera, Universidad Autónoma de Madrid, E-28049 Madrid, Spain.

[*]E-mail: gabino.rubio@uam.es and elsa.prada@uam.es

**Abstract**

Few-layer GaSe is one of the latest additions to the family of 2D semiconducting crystals whose properties under strain are still relatively unexplored. Here, we study rippled nanosheets that exhibit a periodic compressive and tensile strain of up to 5%. The strain profile modifies the local optoelectronic properties of the alternating compressive and tensile regions, which translates into a remarkable shift of the optical absorption band-edge of up to 1.2 eV between crests and valleys. Our experimental observations are supported by theoretical results from density functional theory calculations performed for monolayers and multilayers (up to 7 layers) under tensile and compressive strain. This large band gap tunability can be explained through a combined analysis of the elastic response of Ga atoms to strain and the symmetry of the wave functions.

# 1. Introduction

Strain engineering [1, 2] of two-dimensional (2D) crystals is a technologically promising prospect as well as an attractive field of research in basic science, due to the remarkable tunability of their electronic properties under mechanical deformations, in particular when the number of layers is small. The electrical and optical responses under elastic strain fields are nowadays intensively studied in graphene, transition metal dichalcogenides, and black phosphorus, to name a few cases [3-8]. These 2D crystals belong to the family of layered van der Waals (vdW) materials, characterized by strong intra-layer and relatively weak inter-layer interactions. Remarkably, they can be isolated and thinned down to flakes ranging from tens of layers down to one monolayer [9]. As opposed to their bulk counterparts, these materials in 2D form are highly stretchable, bendable, and even foldable, withstanding strains of up to 10%-25% [3-6]. This translates into, for instance, highly tunable absorption and photoluminescence responses with applications in photovoltaics [10-12], photonics [13-15] and 2D optoelectronics [16]. This remarkable behavior is often accompanied by a strong dependence on the number of layers, adding an extra degree of freedom to create integrated layered materials nanoelectronics.

One of the less explored families of 2D semiconducting crystals is the one formed by post-transition group III monochalcogenides, which share the common formula MX (M = Ga, In and X = S, Se, Te) [17]. A monolayer is formed by two AA-stacked hexagonal sublayers of M atoms sandwiched between two hexagonal sublayers of chalcogen atoms (X) [17, 18]. Here we focus our attention on GaSe. In its bulk form it has a direct gap of 2.1 eV [19, 20], but, as the number of layers decreases, the gap turns quasi-direct, attains a value of 2.4 eV [21-23], and the valence band becomes like an inverted Mexican hat.

Strain engineering is a powerful tool to tune the bandgap of semiconducting 2D materials, and a variety of techniques have been employed to accomplish mechanical deformation: bending of a flexible substrate for $MoS_2$ [24-26] and GaSe [27], substrate elongation for $WS_2$ [28], piezoelectric stretching for $MoS_2$ [29] and controlled rippling for $MoS_2$ [8], $ReSe_2$ [30], $TiS_3$ [31] and GaSe [32].

Here, we investigate rippled flakes down to a 6.3 nm thickness that exhibit a periodic compressive and tensile strain of up to 5%. This strain translates into a shift of the optical absorption band-edge of up to 1.2 eV between crests and valleys, the highest strain-induced energy variation reported to date, and comparable to the variation that can be obtained in quantum confined structures. [23, 33-36] We rationalize these experimental observations with the help of DFT calculations for monolayers and multilayers (up to 7 layers) subjected to strain values similar to those in the experiments, finding excellent agreement for the range of strains where the gap changes linearly and isotropically.

# 2. Methods

*2.1 Micro-transmittance measurements.*

We illuminate the sample using a white light source with enhanced infrared emission (Thorlabs; OSL2FB+OSL2BIR). The sample is actuated by linear positioners with a repeatability of 50 nm. The light transmitted through the flakes is collected by a microscope objective (50X, NA 0.8). A multimode fiber, with a 25 μm diameter core, is placed at the image plane, which corresponds to a 500 nm diameter spot at the focal plane on the sample. The fiber is coupled to a spectrograph equipped with a cooled CCD detector (-60 °C) and spectral resolution of 1 nm (see supplementary information).

*2.2 DFT calculations*

The DFT calculations have been performed with the CRYSTAL14 code [37-39]. The crystal structures were optimized with the PBE functional [40], while the HSE06 hybrid functional [41, 42] was chosen for the presented band structure. A 16×16×1 Monkhorst-Pack k-point mesh was used and all the calculations were converged on the size of the basis set. We ignore the weak spin-orbit interaction [43] in this material since it is not relevant in the large gap variation ranges that we address in this work.

# 3. Results and discussion

*3.1 Experimental measurements*

We obtain highly strained few layer GaSe by mechanical exfoliation of bulk GaSe (HQ Graphene). The micromechanical exfoliation of the flakes involves several steps. First and adhesive tape (Nitto, SPV 224) is used to extract some thick GaSe flakes from the bulk crystal. Next, we repeatedly exfoliate the flakes adhered to the tape using another fresh and clean tape. The remaining flakes are inspected with an optical microscope, and once they are observed to be thin enough they are transferred to a clean and fresh PDMS substrate. [44]

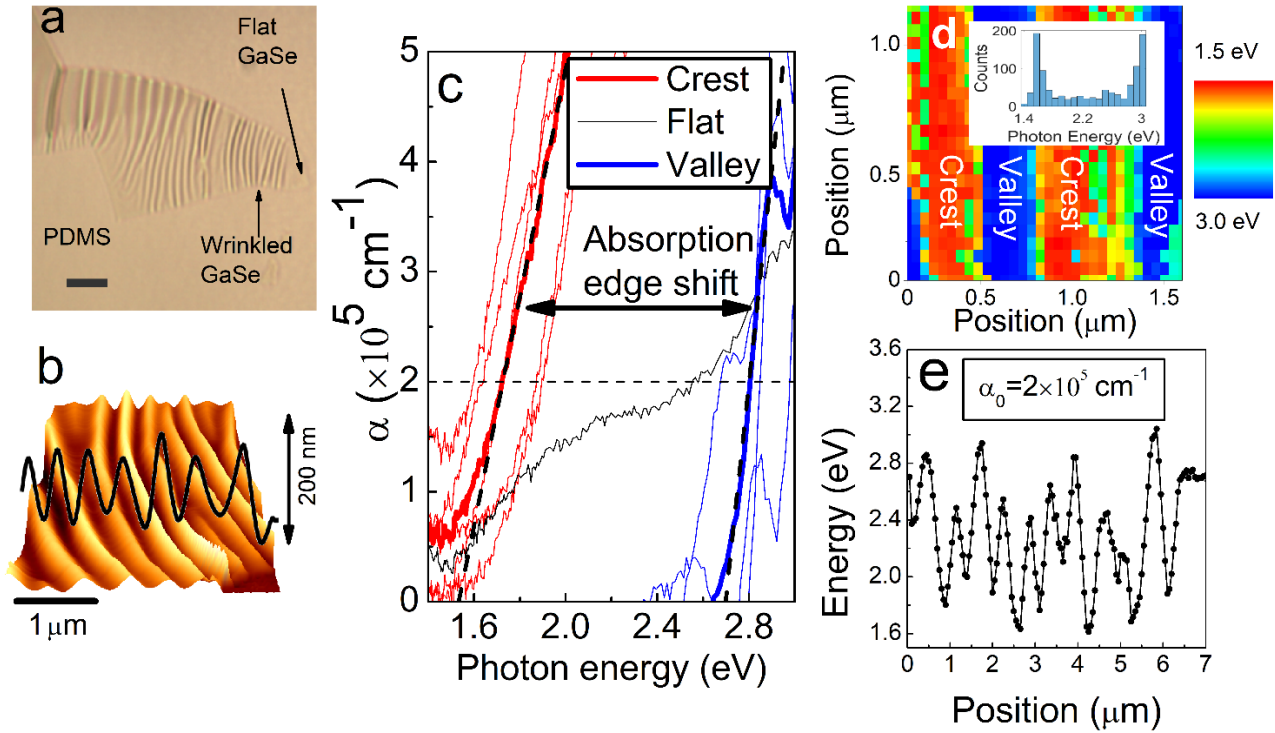

**Figure 1.** Strained few layer semiconducting GaSe crystal. (a) Transmission-mode optical image of a wrinkled 6.3 nm (7 layers) thick GaSe nanosheet on an elastomeric substrate (PDMS). Scale bar is 5 μm. (b) An atomic force topography image of the wrinkled nanosheet, showing a peak-to-peak amplitude of 200 nm and a period of 500-700 nm. (c) Optical absorption spectra at different crests (red solid lines) and valleys (blue solid lines) of the wrinkles and at a flat unstrained region (black solid line). An absorption edge shift of up to ∼1.2 eV is found between crests and valleys. (d) High resolution iso-absorption map of the region of the sample of the energy at which $\alpha_0 = 2 \times 10^5$ cm$^{-1}$. The absorption edge energy variation follows the period of the wrinkles in the direction of the uniaxial compressive strain while being constant along crests and valleys. Inset: Histogram of the iso-absorption energy of the iso-absorption map in (d). A bimodal distribution, which is expected for a sinusoidal strain profile and a linear strain versus band-edge shift relationship (see Supporting Information), is found where the peak at 1.6 eV corresponds to the absorption edge at the tensile regions (crests) and the peak at 3.0 eV corresponds to the compressed regions (valleys). (e) Iso-absorption profile at $\alpha_0$ of the absorption edge energy along ten ripples. The period of the profile is, as for the ripples, ∼ 500-700 nm. The absorption edge energy is blueshifted at compressed valleys while redshifted at the tensile strained crests, reaching a maximum absorption edge shift of up to 1.2 eV.

The optical absorption edge of rippled-strained samples is measured by a scanning micro-transmittance technique [45]. Transmitted light is collected by a 50X 0.8 NA objective, which yields a diffraction limited resolution of 300 nm at 500 nm wavelength. An optical fiber is used as the light detector pinhole and analyzed with a spectrograph equipped with a cooled CCD. We use atomic force microscopy (AFM) to determine the amplitude and the period of the ripples. In order to generate the ripples in the nanosheets, we first prestretch the elastomeric substrate (poly-dimethylsiloxane, PDMS) by bending it and we later deposit the GaSe nanosheets on it. By releasing the substrate, wrinkles are generated in the GaSe flakes if the initial strain exceeds a critical value. This procedure, which is known as buckling-induced rippling [46] generates uniaxial compression and has been successfully used to generate periodic strain profiles in other 2D materials to study their optical properties. [7, 8, 27, 30, 32, 47, 48] The maximum local tensile strain takes place at the crests of the wrinkles, where strain $\varepsilon$ is positive, while at the valleys the compressive strain is negative. The high buckling induced strain relies on the large mismatch between the Young modulus of the nanosheet and the substrate, as well as on the reduced nanosheet thickness [46, 49]. (see Supplementary Information). In addition, this fabrication method yields a fully supported 2D crystal, both at valleys and crests, thus providing a mechanically stable substrate and uniform environment screening.

Figure 1(a) shows a transmission mode optical microscopy image of a 7 layer thick flake on a PDMS substrate where ripples and a flat unstrained region can be distinguished. By measuring the micro-transmittance spectrum at the unstrained region, we determine the thickness of the wrinkled nanosheets. To determine the GaSe thickness, we first fabricate flat samples on PDMS by mechanical exfoliation from bulk GaSe and measure transmittance $T=I/I_0$, where $I$ is the spectrum at the sample and $I_0$ is the spectrum at the elastomeric substrate. Afterwards, we transfer the GaSe flakes onto a Si substrate with a capping layer of 290 nm $SiO_2$ by a dry deterministic transfer method. [50] An AFM was used to measure the thickness of the flakes. Combining both measurements, we relate the transmittance of a flat nanosheet with its number of layers. See supplementary information for more details.

The GaSe ripples have a sinusoidal profile with a period of ~ 500-700 nm, which is expected to be directly proportional to the nanosheet thickness [49]. This is the main reason to select 7 layer flakes: thinner flakes would yield a shorter ripple period, beyond the optical diffraction limit, making local micro-transmittance measurements unfeasible. On the other hand, 7 layer flakes are thin enough to sustain large strain values, as shown below. The amplitude of the wrinkles is obtained by inspecting the topography of the sample with an AFM in contact mode (see figure 1(b)). We observe a peak-to-peak amplitude of 200 nm and a period of ~ 500-700 nm. From these parameters, we estimate that the maximum strain in the 7 layer GaSe nanosheet is $\pm 5\%$ [49].

We have studied the optical properties of the strained GaSe crystal by using a diffraction limited spatially-resolved scanning micro-transmittance setup. In a semiconducting material, the absorption edge can be identified by an increase of the absorption coefficient $\alpha$ above a certain photon energy, due to the excitation of carriers from the valence to the conduction band. We calculate the absorption coefficient from the optical transmittance $T$ and flake thickness $d$ following Lambert law: $\alpha = -\log(T)/d$. In a few layer GaSe flake, the absorption edge energy dramatically shifts at wrinkle crests, valleys and flat regions (see figure 1(c)), that is, at spots with different strain. At the ripple crests, the absorption edge shifts to lower energies due to the closing of the semiconducting gap for tensile strained GaSe. In contrast, at ripple valleys the absorption edge shifts to higher energies. To quantify the shift of the absorption edge at the crests and valleys, we consider a linear fit of the absorption edge as a function of photon energy, and assign the energy at which the fit crosses $\alpha = 0$. At ripple crests, where $\varepsilon > 0$, the absorption edge plummets to a value of 1.5 eV and at ripple valleys, where $\varepsilon < 0$, the absorption edge reaches 2.7 eV, which gives an overall change of ~ 1.2 eV. To our knowledge, this is the highest strain induced band gap energy shift reported in 2D semiconducting post-transition metal monochalcogenides. The absorption edge energy shift can be obtained building an iso-absorption map or profile, that is, a map or profile of the energy at which the absorption coefficient at each position along the ripples reaches a certain value. These iso-absorption maps follow the periodic pattern of the ripples with alternating tensile and compressive strain at crests and valleys, respectively. A high resolution iso-absorption map of two wrinkles is shown in figure 1(d). This map shows the energy at which absorption coefficient is $\alpha_0 = 2 \times 10^5$ cm$^{-1}$ and these energies follow a periodic pattern in the direction where we have applied the compressive strain and are constant along the crests and valleys. Figure 1(e) shows the iso-absorption profile at $\alpha_0$ along ten consecutives ripples. There is an oscillation of the absorption edge energy corresponding to alternating valleys and crests of the ripples.

Previous attempts [33, 51-53] to measure micro-photoluminescence (PL) in few layer GaSe have been proven to be elusive because PL intensity is strongly quenched and indistinguishable from the noise (see figure S4(b), Supplementary Information). In addition, it has been reported that the light power density required for PL measurements induces degradation in the sample by thermal oxidation [33, 51-54]. Alternatively, optical absorption requires a power density four orders of magnitude lower. This leaves low power density absorption edge measurements as the sole straightforward experimental method to determine the optical gap of the thinnest GaSe flakes (more details in Supplementary Information).

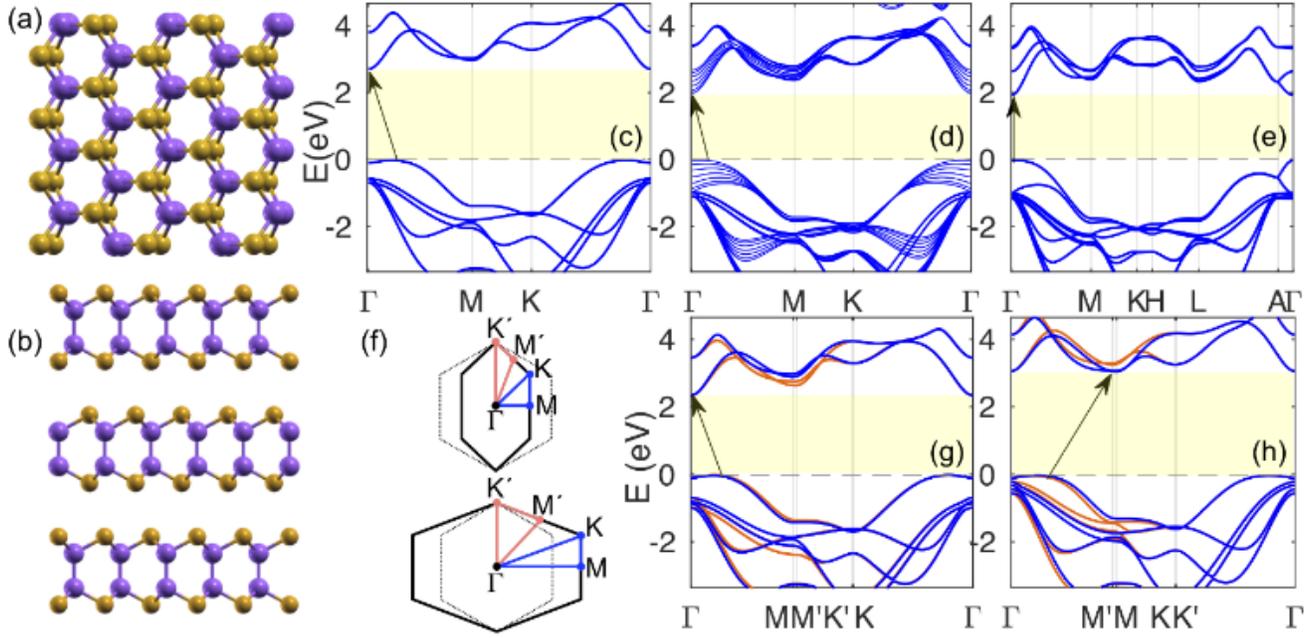

**Figure 2**. (a) Top view (tilted) of monolayer GaSe with Ga atoms shown in purple and Se in yellow. (b) Lateral view of AB-stacked ε-GaSe. DFT band structure for (c) a monolayer, (d) seven layers and (e) bulk. (f) Schematic first BZ with uniaxial tensile (top) and compressive (bottom) strains along the AC direction. The BZ without strain is also shown as a reference. Two different paths between high symmetry points for the deformed BZ are highlighted in blue and orange. Monolayer band structure with 3% tensile (g) and -3% compressive (h) strains, both in the AC direction. Blue and orange bands correspond to the two paths shown in (f). Arrows indicate the energy gap in all cases.

*3.2 Theoretical calculations*

The crystal structure of GaSe is presented in figure 2(a-b). Multilayers form the most common non-centrosymmetric AB stacking polytype, known as ε-GaSe. In figure 2(c-e) we show the band structure of single-layer, 7 layer and bulk GaSe as obtained from DFT. As previously reported in the literature [22, 55, 56], we find that monolayer GaSe is a quasi-direct band gap semiconductor with gap energy $E_g$=2.73 eV. The conduction-band (CB) minimum is located at the Γ point while the valence-band (VB) maximum is slightly shifted towards the M-point (Figure 2(c)). The VB dispersion near Γ has an inverted Mexican hat shape, which is present in all three gallium chalcogenides [17, 18, 57]. By increasing the number of layers, the gap decreases and the Mexican hat band shape flattens, although is still visible for 7 layers (see figure 2(d)). The VB eventually acquires a typical inverted parabolic shape around Γ for the bulk crystal, as shown in figure 2(e), where we obtain a gap of 1.95 eV, in agreement with several experiments which report a value around 2.0-2.1 eV [19, 20]. In addition, the band structure has also an impact on the absorption coefficient. The density of states associated with the valence band is comparatively much higher at the band edge in the case of few-layers than in bulk. This is due to the almost flat band structure for a number of layers around or below 10. [17] Thus one can expect a higher absorption coefficient for our flakes than for the bulk material where the valence band presents a standard curvature. The experimental method used here is suitable to study band structure modifications for which the strain shifts the band gap energy vertically, corresponding to optical transitions.

We first consider the effect of uniaxial tensile and compressive strains applied in the armchair (AC) direction for a monolayer. The main effect of tensile strain is that of decreasing the gap by shifting the CB minimum downwards at the Γ point (with respect to the M point) and splitting the highest VB further from the rest (see figure 2(g)). This behavior can be rationalized as follows. The CB and VB at Γ belong to different irreducible representations (IRs) of the symmetry group of the crystal. Following the notation of ref [18], the VB belongs to $Γ_1^+$ and the CB belongs to $Γ_2^-$, hence they transform differently under an in-plane mirror operation. Although they are not a simple bonding-antibonding pair with respect to the mirror plane (they have a different in-plane nodal structure [18]), still the $sp_z$ orbitals of Ga atoms dominate the splitting (the gap), which is expected to increase with an increasing hybridization of these and vice versa. Our calculations reveal, somewhat counter-intuitively according to standard elasticity theory, that the vertical distance between Ga atoms in the same primitive cell increases with tensile strain (the opposite with compressive strain), decreasing the hybridization and thus reducing the gap. Although the change in bond length between Ga atoms is small (up to 2%), it is enough to have a visible effect due to the directional character of the $sp_z$ orbitals. The reason behind this unexpected elastic behavior of the Ga atoms can be traced back to the ionic character of the Ga-Se bond. Under tensile strain, the monolayer thickness is decreased. This, in turn, reduces the Ga-Se bond length and the bond becomes more ionic. This withdraws charge from the Ga atom, thus reducing the bond between Ga atoms which separate from each other.

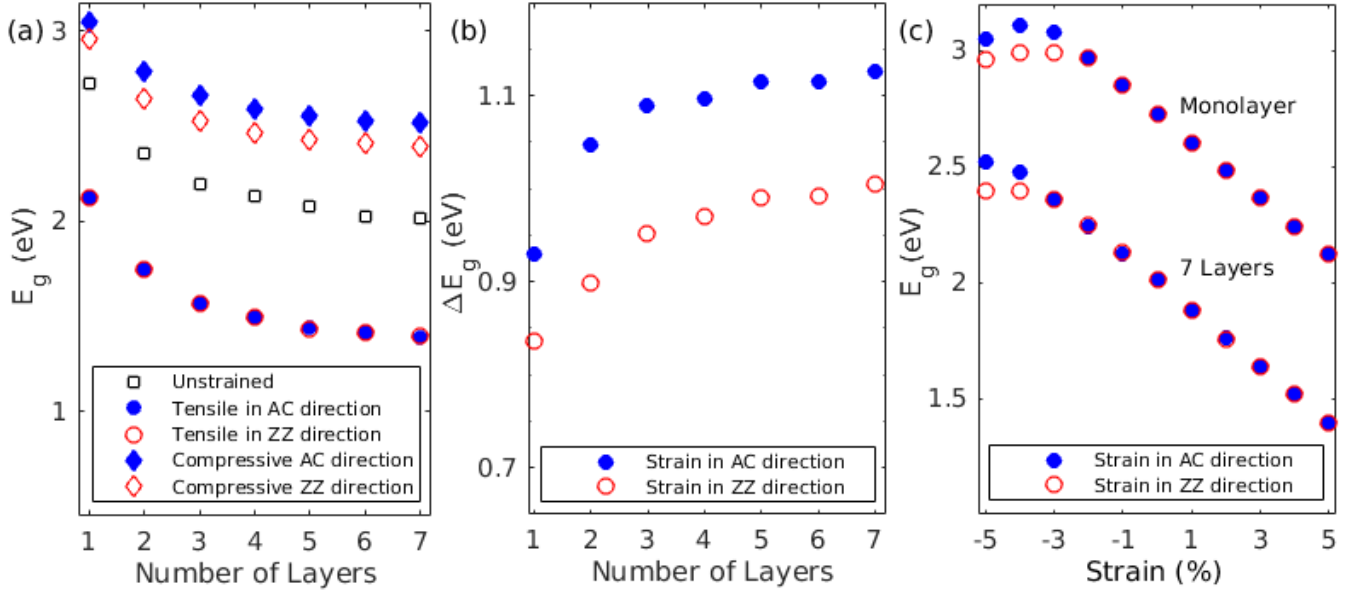

**Figure 3**. Theoretical DFT analysis of the bandgap. (a) Calculated energy gap of ε-GaSe as a function of the number of layers, N, for the compressed (-5%), unstrained and tensile strained (+5%) cases. Uniaxial strain in the AC (ZZ) direction is represented with solid blue (empty red) symbols, while the relaxed crystal is represented in black. (b) Calculated energy gap difference $\Delta E_g$ between ±5% strain for both crystal directions as a function of N. (c) Calculated energy gap as a function of strain for monolayer and 7 layer crystals. Note that we only include results for integer values of strain in %. The linear behavior of $E_g$ with strain is lost for compressive strains larger than -2% for monolayer and -3% for N=7, signalling the strain for which the gap changes from quasi-direct to indirect.

Conversely, when the crystal is compressed, the Ga-Ga distance reduces and the gap increases remaining quasi-direct until up to -2%. Above this value the CB minimum at the Γ-point crosses over to the M-point and the gap turns to indirect (see figure 2(h)). This transition can be understood by examining the orbital nature of the CB at the M point. Here the s character of the wave function increases compared to that at Γ and this makes it less sensitive to changes in the Ga-Ga distance. The shape of the VB does not change until ∼-5% strain, when the lower energy VBs take over and the VB maximum shifts to the Γ point (not shown). Again, this can be understood by the fact that the lower VBs belong to IRs $\Gamma_3^{-/+}$ which transform according to $p_x$, $p_y$ orbitals and are much less sensitive to changes in distances along the z direction.

In figure 3 we show a theoretical DFT analysis of the behavior of the gap with strain and the number of layers. As can be seen in figure 3(a), the gap decreases with the number of layers and it does so in a similar fashion for the unstrained, the tensile and the compressive strained flakes.

In figure 3(b) we show the energy gap difference between ±5% strain for both crystal directions as a function of N. For 7 layers, which is the thickness of our flakes in the experiment, we get $\Delta E_g \approx 1$ eV for strain in the ZZ direction and $\Delta E_g \approx 1.1$ eV for strain in the AC one, to be compared to the experimental value $\Delta E_g \approx 1.2$ eV. The differences found between the AC and ZZ cases do not seem to be significant compared to the large band gap variation so as to consider this variable when contrasting theory and experimental results, which are, overall, in very good agreement.

In figure 3 (c) we can observe the behavior of the gap with the applied strain in %, both for monolayer as well as for 7 layers. Eg changes linearly with strain for almost all the strain range considered. In this range, the gap remains direct and its variation is moreover isotropic with strain direction. However, this linear and isotropic behavior is lost for compressive strain values larger than -2% for monolayer and -3% for N=7, where the gap changes from quasi-direct to indirect (and becomes direction dependent). The absorption measurements can capture the variation of the gap with strain in the linear region where the gap is quasi-direct, corresponding to optical transitions, while it may not capture the onset of the indirect gap for the (small) region of compressive strains where it appears. The theoretical variation of the gap around the Γ point considering the whole strain range from +5% to -5% results in a slightly larger isotropic gap variation $\Delta E_g \approx 1.2$ eV.

Note that in this analysis the possible presence of excitons and their associated absorption gap reduction has been ignored. This assumption is justified since Budweg et al. [58] demonstrated the absence of excitons for GaSe flakes thinner than 8 layers, both experimentally and theoretically. Based on these theoretical results, we conclude that the response of few-layer GaSe to strain is isotropic in the entire quasi-direct band gap regime. Moreover, in this region we find a strain coefficient of 124 meV/% for N=7 and of 100 meV/% for monolayer. A comparative table of the strain coefficients for several 2D crystals can be found in ref [59]. According to it, the strain coefficient that we find for GaSe is one of the highest strain coefficients reported in the literature, much larger than those for dichalcogenides, of the order of BP's and slightly smaller than InSe's. Previous reports also find a high strain coefficient for thicker GaSe flakes, [27] which do not allow for such high strain as in our case, hence showing a significantly smaller absorption edge energy shift.

## 4. Conclusion

In summary, we find a strong modulation of the optical gap of strain-engineered few layer GaSe. A periodic compressive and tensile strain profile of up to 5% is obtained by buckling induced rippling of the GaSe nanosheets. We locally measure diffraction-limited micro-transmittance to obtain the optical absorption band-edge and find a blueshift for compressive strain and a redshift for tensile strain. Our DFT calculations show that the strong band gap variation is due to the applied in-plane stress. In the range of 5% tensile to -3% compressive deformations, wherein the electronic response is isotropic, we predict a linear change of the gap energy resulting in a strain coefficient of 124 meV/% for a 7 layer crystal, in agreement with our experimental findings. The large strain attained in the experiments yields an absorption edge shift as large as 1.2 eV from tensile to compressive strain, which is, to our knowledge, the highest strain-induced band gap energy variation in 2D semiconducting crystals reported to date.

**ORCID iDs**

David Maeso: 0000-0001-5105-1789

Sahar Pakdel: 0000-0002-4676-0780

Hernán Santos: 0000-0002-6130-2163

Nicolás Agraït: 0000-0003-4840-5851

Juan José Palacios: 0000-0003-2378-0866

Elsa Prada: 0000-0001-7522-4795

Gabino Rubio-Bollinger: 0000-0001-7864-8980

**Acknowledgements**

Research supported by the Spanish MINECO through Grants MAT2017-88693-R, MAT2014-57915-R, FIS2016-80434- P (AEI/FEDER, EU), BES-2015-071316, the Ramón y Cajal programme RYC-2011-09345, the Fundación Ramón Areces and the María de Maeztu Programme for Units of Excellence in R&D (MDM-2014-0377), as well as from the Comunidad Autónoma de Madrid (CAM) MAD2D-CM Program (S2013/MIT-3007) and the European Union Seventh Framework Programme under grant agreement No. 604391 Graphene Flagship. We acknowledge the computer resources and assistance provided by the Centro de Computación Científica of the Universidad Autónoma de Madrid.

# Supplementary Information

# Strong Modulation of Optical Properties in Rippled 2D GaSe *via* Strain Engineering

**David Maeso[1], Sahar Pakdel[1,2], Hernán Santos[1], Nicolás Agraït[1,3,4,5], Juan José Palacios[1,4,5], Elsa Prada[1,4,5*] and Gabino Rubio-Bollinger[1,4,5*]**

[1]Departamento de Física de la Materia Condensada, Universidad Autónoma de Madrid, E-28049 Madrid, Spain.

[2]School of Electrical and Computer Engineering, University College of Engineering, University of Tehran, Tehran 14395-515, Iran.

[3]Instituto Madrileño de Estudios Avanzados en Nanociencia (IMDEA-Nanociencia), E-28049 Madrid, Spain.

[4]Condensed Matter Physics Center (IFIMAC), Universidad Autónoma de Madrid, E-28049 Madrid, Spain.

[5]Instituto Nicolás Cabrera, Universidad Autónoma de Madrid, E-28049 Madrid, Spain.

[*]E-mail: gabino.rubio@uam.es and elsa.prada@uam.es

Table of contents:

### 1. Buckling-induced rippling

We fabricate the strained GaSe nanosheets by using an elastomeric substrate (PDMS) whose Young's modulus is substantially smaller than the GaSe Young modulus [1-5]. We prestretch the PDMS substrate by bending it and deposit few-layer GaSe nanosheets on its convex surface, using the mechanical exfoliation method, from GaSe bulk (HQgraphene). We release the substrate, which becomes flat again, thus compressing the GaSe flakes on its surface, producing wrinkles due to buckling-induced rippling caused by the large mismatch between the elastic properties of the nanosheet and the elastomeric substrate.

The rippled pattern usually follows the direction of the compressive stress, but due to the relative GaSe flakes crystalline orientation, in some cases the pattern of the ripples is irregular or even absent. It is only after inspection with an optical microscope that some flakes which show a regular rippled pattern are selected.

Thin film wrinkling can be modelled as a force balance between the stiff thin film and the soft elastic substrate [6-9]:

$$\bar{E}_f I \frac{d^4 y}{dx^4} + F \frac{d^2 y}{dx^2} + ky = 0 \tag{S1}$$

where $\bar{E} = \frac{E}{(1-\nu^2)}$ is the plane-strain modulus, E is the Young's modulus, $\nu$ the Poisson ratio, I=wh³/12 is the moment of inertia, w and h are the width and the thickness of the thin film, F the applied force and $k = \bar{E}_s w \pi / \lambda$ is the Winkler's modulus of an elastic half-space [8]. Subscripts $f$ and $s$ denote film and substrate, respectively. The period of the wrinkles is:

$$\lambda = 2\pi h \left(\frac{\bar{E}_f}{3\bar{E}_s}\right)^{1/3} \tag{S2}$$

and the critical amount of stress to wrinkle the thin film is given by:

$$\sigma_c = \frac{F_c}{hw} = \left(\frac{9}{64} \bar{E}_f \bar{E}_s^2\right)^{1/3} \tag{S3}$$

Therefore the critical strain is:

$$\varepsilon_c = \frac{\sigma_c}{\bar{E}_f} = \frac{1}{4}\left(\frac{3\bar{E}_s}{\bar{E}_f}\right)^{2/3} \tag{S4}$$

Wrinkle amplitude is expressed as:

$$A = h\sqrt{\frac{\varepsilon}{\varepsilon_c} - 1} \tag{S5}$$

One can estimate the strain in the wrinkles:

$$\varepsilon = \varepsilon_c \left(1 + \frac{A^2}{h^2}\right) = \frac{1}{4}\left(\frac{3\bar{E}_s}{\bar{E}_f}\right)^{2/3} \left(1 + \frac{A^2}{h^2}\right). \tag{S6}$$

We inspect the topography of the wrinkles and measure their amplitude using an atomic force microscopy (AFM) in topography contact mode. To study the flake thickness we combine micro-transmittance measurements with AFM measurements (see GaSe thickness determination).

Considering the PDMS Young modulus $E_s = 500$ kPa and its Poisson ratio $\nu_{PDMS} = 0.5$ [1], the GaSe Young modulus $E_f = 92.3$ GPa and its Poisson ratio $\nu_f = 0.24$ [2], the GaSe thickness $h = 6.3$ nm and amplitude $A = 100$ nm, one finds, using equation S6 that the maximum strain in the ripples (both compressive and tensile) is $\varepsilon_{max} = 5\%$.

## 2. GaSe thickness determination

We first fabricate flat samples on PDMS by mechanical exfoliation from bulk GaSe. We measure the transmittance $T=I/I_0$, where $I$ is the spectrum at the sample and $I_0$ is the spectrum at the elastomeric substrate, at flat GaSe nanosheets with thicknesses from 7 to 30L at three different wavelengths (500, 650 and 800 nm) as shown in figure S1(a). Afterwards, we transfer the GaSe flakes onto a Si substrate with a capping layer of 290 nm $SiO_2$ (see figure S1(b)) by a dry deterministic transfer method [10]. An AFM was used to measure the thickness of the transferred flakes in contact mode, to avoid possible artefacts, as shown in figure S1(c). Combining both measurements, we relate the transmission of a flat nanosheet with its number of layers (see figure S1(d)). By comparing transmission at those wavelengths in a flat region of the strained GaSe nanosheet with figure S1(d) we determine its flake thickness.

Figure S1(d) shows, for the thinnest flakes, a transmission very close to one, suggesting negligible reflectivity. However, strictly, the figure shows the quotient between the transmission of the flake and the bare substrate ($I/I_0$). Because both interfaces yield a similar reflectivity for the illumination beam, the quotient tends to hide the true value of the reflectivity, which cannot be straightforwardly obtained for the sole measurement of ($I/I_0$).

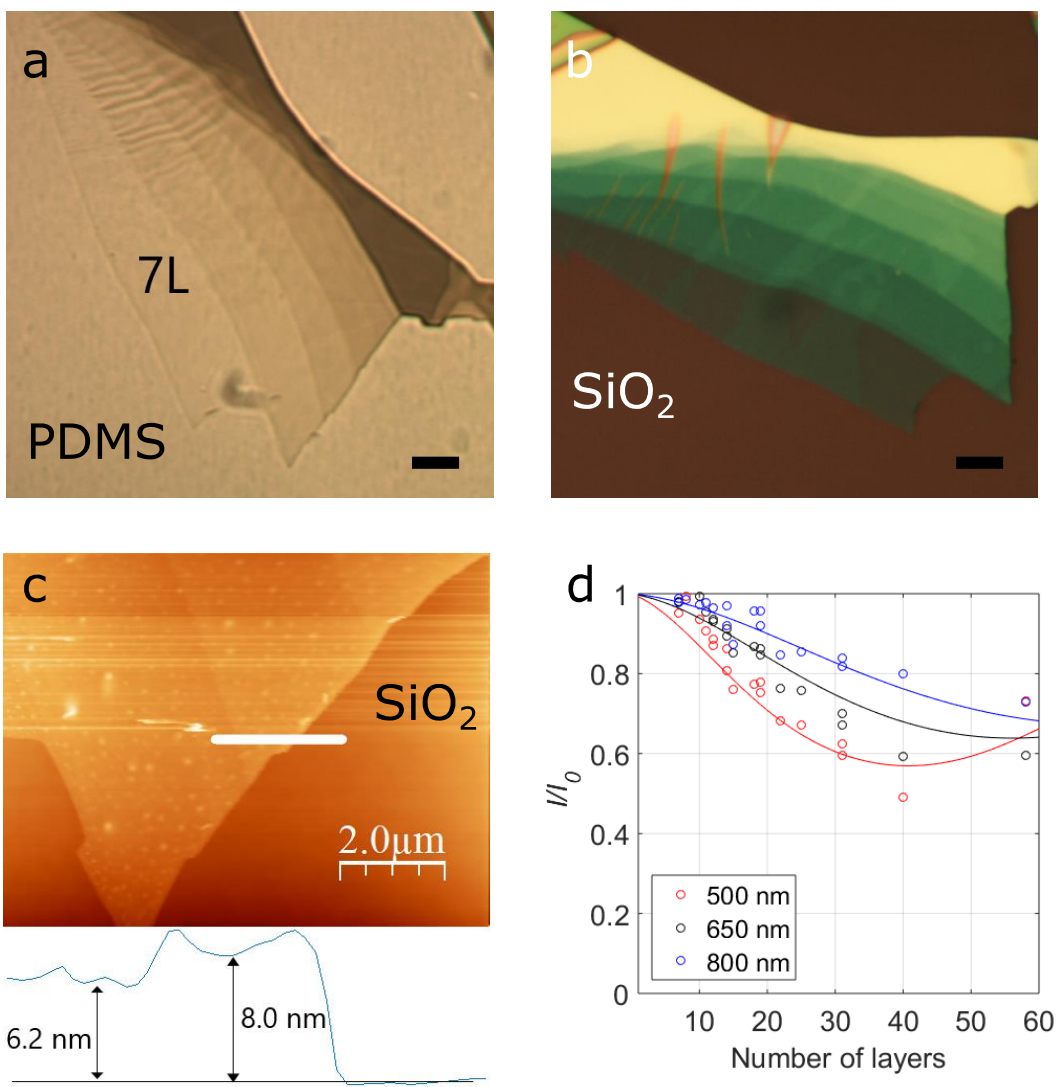

**Figure S1**: Thickness determination of GaSe nanosheets. (a) Transmission optical image of a GaSe flake on an elastomeric substrate (PDMS) where the thinner region is 7 layers. Note that the thinnest region is almost transparent $(T\sim 0.97)$. Scale bar is 5 µm. (b) Reflection optical image of the same GaSe flake but transferred onto 290 nm of SiO$_2$/Si. The optical contrast when the nanosheet is on SiO$_2$/Si is enhanced by the optical interference between the reflection paths at the GaSe/SiO$_2$/Si stacking and substrate. Scale bar is 5 µm. (c) AFM image of a region of the GaSe flake. Inset: Topography profile acquired along the white solid line. (d) $I/I_0$ vs. number of layers at three wavelengths for GaSe thin films. Experimental data (circles) are fitted to the Fresnel equation (solid lines) in order to have a simple and efficient way to obtain the GaSe thickness comparing these fits with the transmittance at the unstrained region of a rippled GaSe nanosheet.

## 3. Experimental Setup

A sketch of the experimental setup is shown in figure S2. We illuminate the sample using an enhanced infrared white halogen light source (OSL2, OSL2BIR, Thorlabs). Two silver coated 90° off-axis parabolic mirrors (MPD00M9-P01 and MPD019-P01, Thorlabs) are used to project the end of the light source fiber on the sample. We use an optical microscope (Nikon Eclipse LV100) and the sample is actuated by two nanoprecise linear positioners (ECSx3030, Attocube) with a positioning repeatability of 50 nm. This stage allows us to scan a selected region of the sample. The light transmitted through the flakes is collected by a microscope objective (50X, NA 0.8, Nikon). Spatial resolution $r$ is diffraction limited by $r = \lambda/2\,NA$ where $\lambda$ is the illumination wavelength and NA is the numerical aperture of the objective. In our system $r \sim 300$ nm for $\lambda = 500$ nm. A 30:70 plate beamsplitter (EBPS1, Thorlabs) divides the transmitted light into two beams. In one of the paths, a CMOS camera (EO-5012C, IDS) is used to inspect the sample. In the other path, we place a multimode fiber, with a 25 µm diameter core, at the image plane, which corresponds to a 500 nm diameter spot at the focal plane on the sample. This fiber is coupled to a single grating Czerny-Turner spectrograph (Shamrock, Andor) with 193 mm focal length and a reflective 300 grooves/mm diffraction grating blazed at a wavelength 500 nm (GR50-0305, Thorlabs). The spectrograph is equipped with a cooled CCD detector with an array of 2000 × 256 pixels and 15 µm pixel size at -60 °C (iDus 416 CCD, Andor).

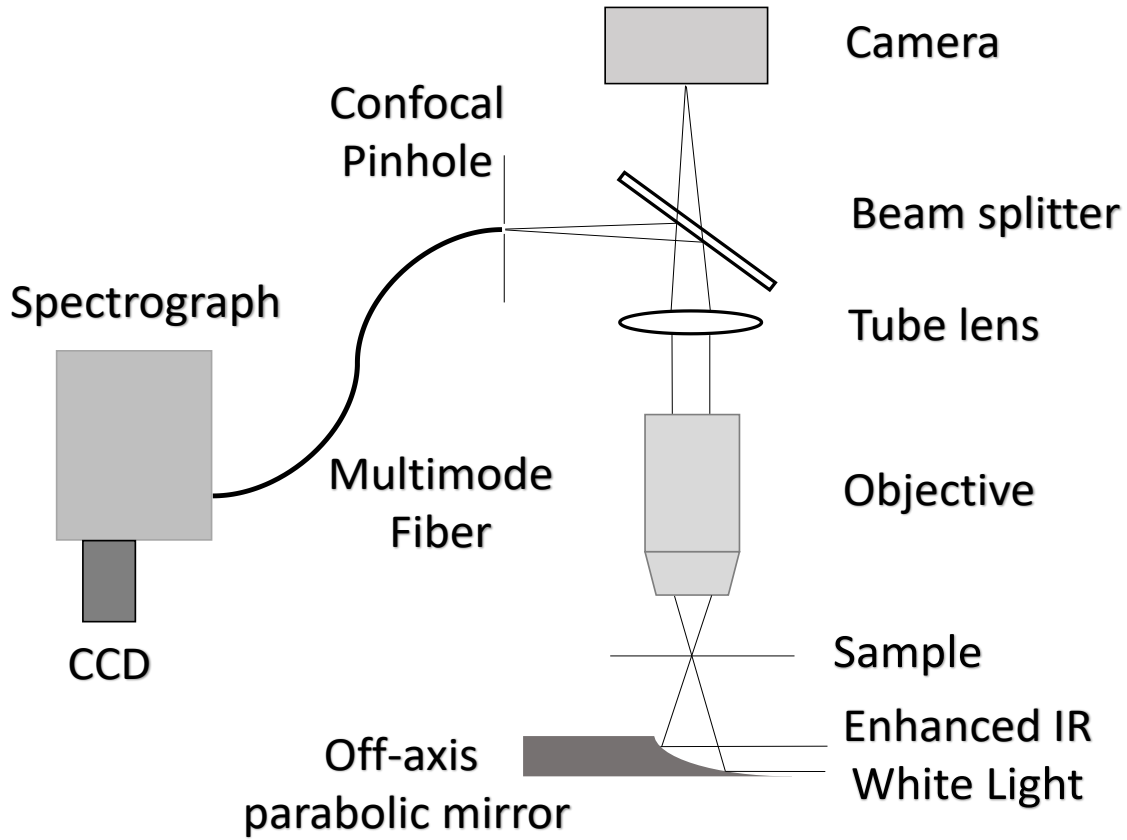

**Figure S2**: Sketch of the micro-transmittance setup to measure micro-absorbance of few-layer GaSe nanosheets.

## 4. Raman and Photoluminescence spectroscopy

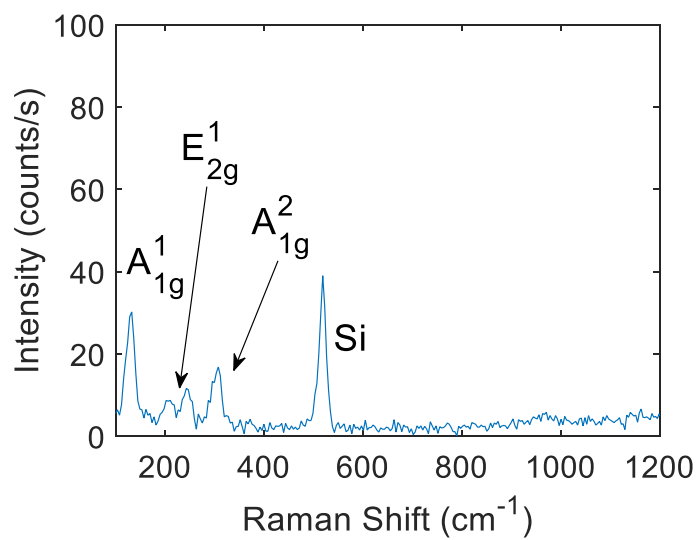

**Figure S3.** Raman spectra of bulk GaSe exfoliated on 290 nm $SiO_2$/Si substrate

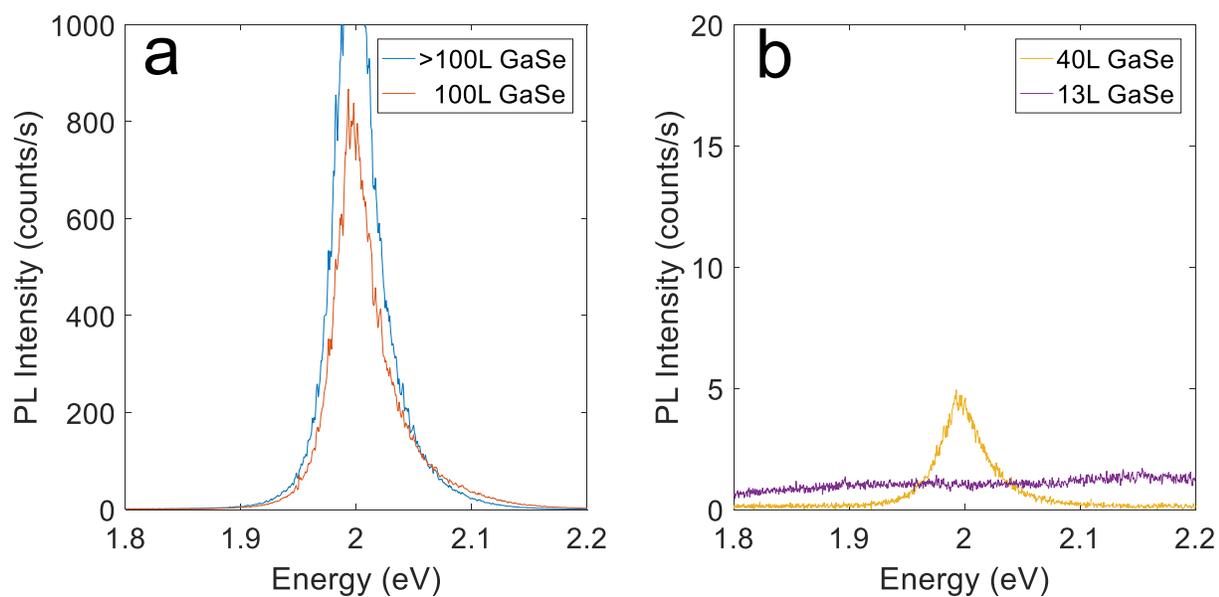

**Figure S4.** Micro-photoluminescence spectrum of GaSe flakes. (a) Photoluminescence of thick GaSe from two regions of the flake showed in figure S1(**Error! Reference source not found.**b). The PL peaks are centered at the bulk band gap energy of ~ 2 eV. (b) Photoluminescence of thin GaSe nanosheet from two thin regions of the flake showed in figure S1(**Error! Reference source not found.**b). The photoluminescence peak height decreases while we reduce the number of layers.

Thin GaSe nanosheets have been characterized by micro-Raman spectroscopy in previous reports and important features have been reported for micro-Raman measurements at room temperature and under ambient conditions: oxidation, laser damage and low photoluminescence response in thin samples [11-15]. Figure S3 shows Raman spectrum of a bulk GaSe flake on 290 nm $SiO_2$/Si. A

532 nm continuous wave laser (Spectra-Physics, Excelsior One), a diffraction limited confocal optical microscope (Nikon Eclipse LV100) and a cooled CCD coupled to a spectrograph (Andor) were used for micro-Raman measurements. The total power at the sample was ~ 100 µW to prevent overheating and laser damage of the samples and the spot of the excitation laser was 1 µm in diameter. We can identify the silicon peak around 520 cm$^{-1}$ and the GaSe peaks $A_{1g}^1, E_{2g}^1$ and $A_{1g}^2$ around 130, 220 and 310 cm$^{-1}$ [11, 12, 16].

Micro-photoluminescence spectroscopy of thick flakes shows a prominent photoluminescence peak around 2.0 eV which is related to the band gap (see figure S4(a)). When we reduce the number of layers to few-layer nanosheets, the photoluminescence peak cannot be distinguished from noise (see figure S4(b)).

5. Iso-absorption Energy Histogram

We show in figure S5 the iso-absorption photon energies at which the absorption coefficient is $\alpha_0 = 2 \times 10^5$ cm$^{-1}$ and the corresponding histogram. We observe that the distribution of the iso-absorption energies follows a bimodal distribution, which is expected for a sinusoidal strain profile and a linear strain versus band-edge shift relationship. The probability distribution is proportional to the absolute value of the inverse of the derivative. In this ideal case, the bimodal distribution would be proportional to $\left|\cos\left(\frac{\pi}{2}\frac{\varepsilon}{\varepsilon_{max}}\right)\right|^{-1}$ where $\varepsilon$ is the strain and $\varepsilon_{max}$ is the absolute value of the maximum strain. Ignoring the divergences at $\varepsilon = \pm\varepsilon_{max}$ the red line in figure S5(b) shows the corresponding band-edge energy distribution. The first peak at 1.6 eV corresponds to the absorption edge at the stretched regions and the second peak at 3.0 eV corresponds to the compressed regions. A remarkable absorption edge shift of 1.4 eV is observed in the histogram.

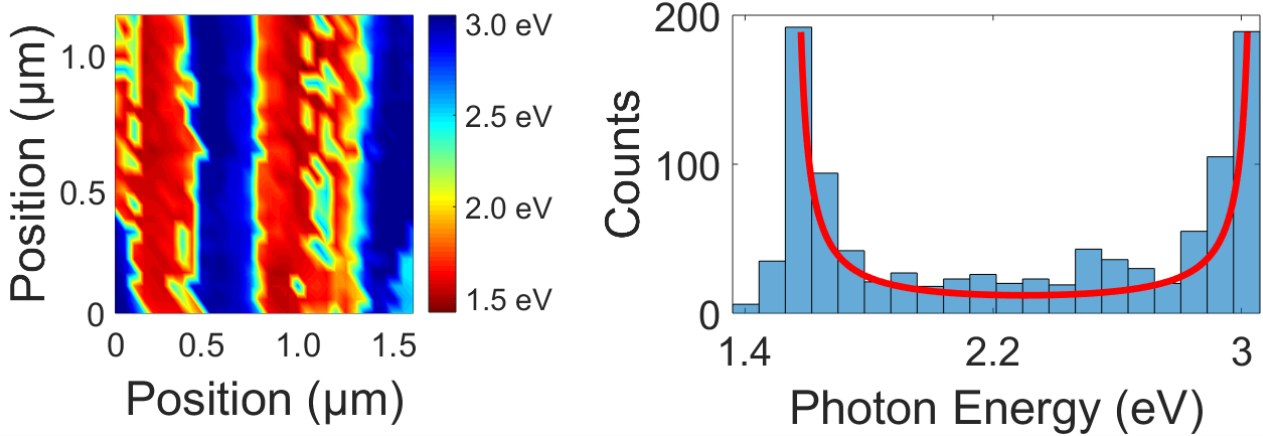

**Figure S1.** (a) High resolution iso-absorption map of the energies at $\alpha_0 = 2 \times 10^5\ cm^{-1}$ of two ripples of GaSe from the main text. (b) Histogram of the iso-absorption energies of the iso-absorption map.

## 6. Ab-initio band structure of 7 layer GaSe with strain

We here provide the band structure of a 7 layer GaSe crystal under ±1%, ±3% and ±5% in-plane strains applied in the AC (Figure S6) and the ZZ direction (Figure S7). These DFT calculations have been performed using HSE functionals as described in the main text.

The response to strain in both directions is the same, except for applied deformations where there is a transition from quasi-direct to indirect gap, which occurs for compressive strains beyond ∼ -5% in the AC case and slightly sooner in the ZZ one.

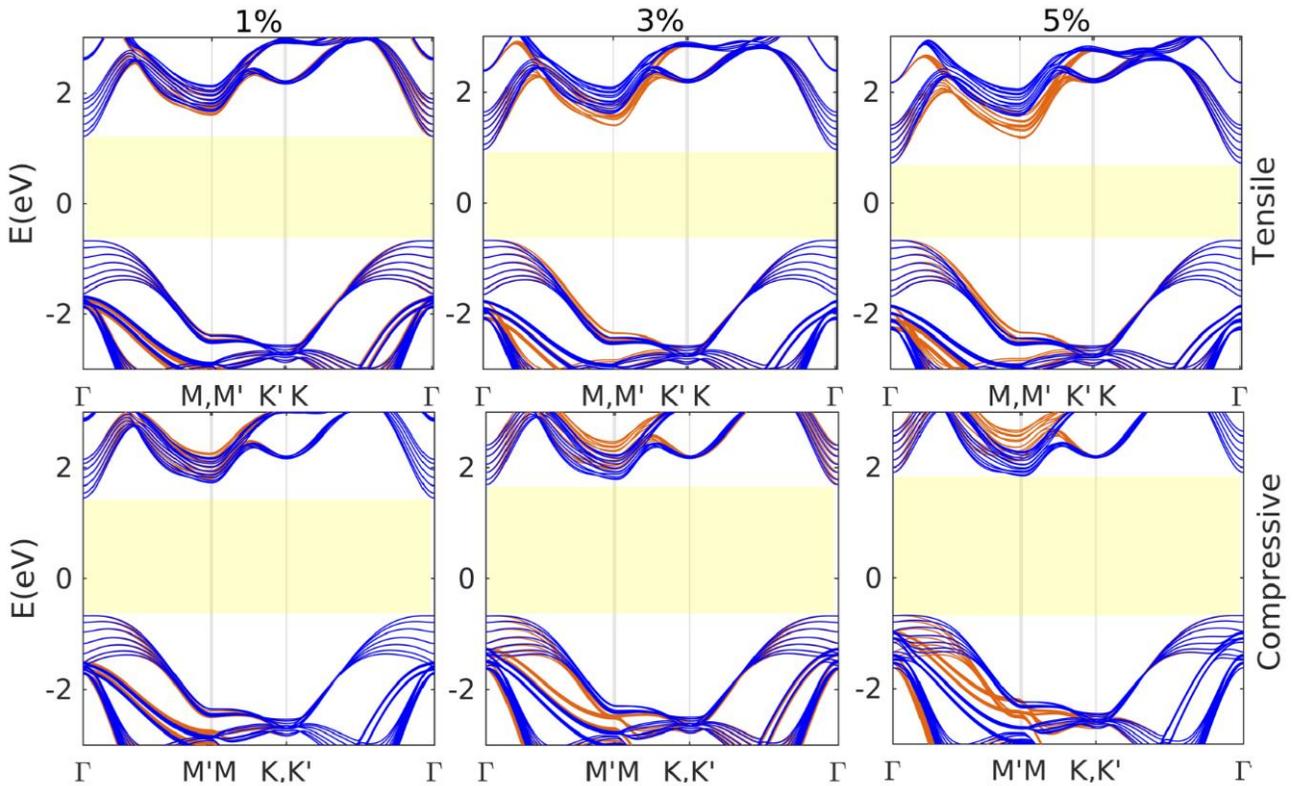

**Figure S6**. Ab-initio band structures for a 7 layer GaSe crystal for different in-plane strains in the AC direction. Upper panels correspond to 1%, 3% and 5% tensile strain, while lower panels correspond to -1%, -3% and -5% compressive strain. Blue (dark) and orange (light) colors correspond to different paths between high symmetry points in the deformed BZ of figure 2(f).

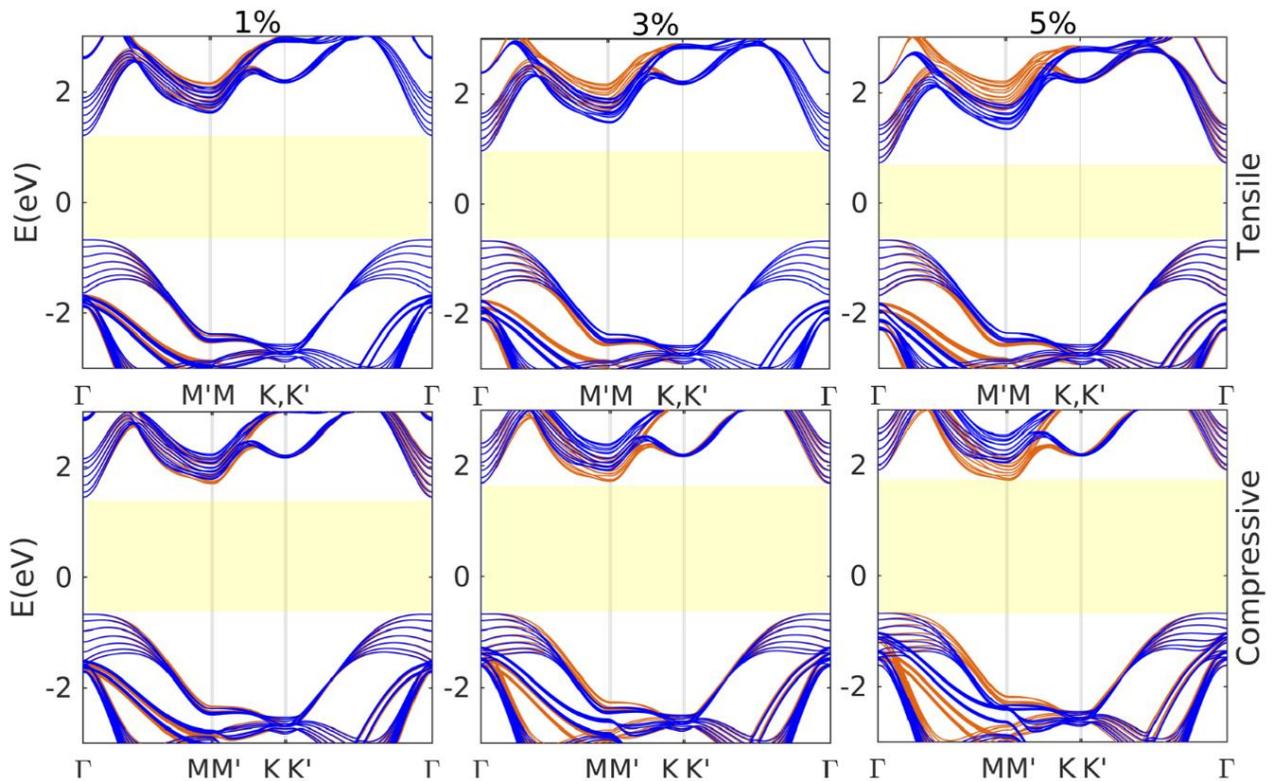

**Figure S7**. Ab-initio band structures for a 7 layer GaSe crystal for different in-plane strains in the ZZ direction. Upper panels correspond to 1%, 3% and 5% tensile strain, while lower panels correspond to -1%, -3% and -5% compressive strain. Blue (dark) and orange (light) colors correspond to different paths between high symmetry points in the deformed BZ of figure 2(f).